\documentclass[%
 reprint,
 amsmath,amssymb,
apl,
tightenlines]{revtex4-1}

\usepackage{graphicx}
\usepackage{dcolumn}
\usepackage{bm}
\usepackage{color}
\usepackage{graphicx}
\usepackage{float}
\usepackage{amsmath}
\usepackage{subfig}
\graphicspath{ {images/} }

\newcommand{\rvec}{\mathbf{r}} 
\newcommand{\Rvec}{\mathbf{R}} 
\newcommand{\rev}[1]{{\color{black} #1}}
\newcommand{\revDK}[1]{{\color{black} #1}}
\newcommand{\etal}{\emph{et al.}}

\begin{document}

\title{Constrained Density Functional Theory Calculation  with Iterative Optimization}


\author{Daniel Kidd}
\affiliation{Department of Physics and Astronomy, Vanderbilt University, Nashville, Tennessee 37235, USA}
\author{A. S. Umar}
\affiliation{Department of Physics and Astronomy, Vanderbilt University, Nashville, Tennessee 37235, USA}
\author{K\'alm\'an Varga}
\email{kalman.varga@vanderbilt.edu}
\affiliation{Department of Physics and Astronomy, Vanderbilt University, Nashville, Tennessee 37235, USA}

\date{\today}

\begin{abstract}
An iterative optimization approach  that  simultaneously minimizes the  energy
and optimizes the Lagrange multipliers enforcing desired
constraints is presented. The method is tested on previously established
benchmark systems and it is proved to be efficient and accurate. 
The approach can also be efficiently used when the
constraint is not a scalar quantity but a spatially varying function
like the charge density distribution.
\end{abstract}

\maketitle



\section{\label{sec:intro}Introduction}

Density functional theory (DFT) \cite{PhysRev.136.B864} is one of the
most important approaches to calculating ground state properties in
molecules and solids. The
extension of DFT for ground state calculations in constrained systems
(cDFT) \cite{dederichs1984ground,PhysRevB.22.5777} opened a new venue
for the description of charge excitations \cite{PhysRevLett.56.2407},
magnetic transitions \cite{PhysRevB.44.13319}, spin dynamics
\cite{doi:10.1063/1.370494} and electron transfer
\cite{doi:10.1063/1.480117}. This technique became a more powerful 
tool with a greatly enhanced range of applicability
through the introduction of a self-consistent formulation
 by Wu and Van Voorhis
\cite{PhysRevA.72.024502}. It is now
implemented in many computer codes using localized basis sets
(NWChem \cite{PhysRevA.72.024502}, QChem \cite{B912085H,doi:10.1063/1.2800022}, SIESTA
\cite{PhysRevB.88.165112}, deMon2k \cite{doi:10.1021/ct200570u}, ADF
\cite{C6CP00528D}), plane waves (CPMD \cite{doi:10.1063/1.3190169},
QuantumESPRESSO \cite{doi:10.1063/1.3326226}, VASP
\cite{PhysRevB.91.054420}),
density matrices (CONQUEST \cite{doi:10.1021/ct100601n}), wavelets 
(BigDFT \cite{doi:10.1021/acs.jctc.5b00057}), and projector augmented
wave (PAW) methods \cite{acs.jctc.6b00815}. 

Armed with these powerful computational tools the cDFT has been
intensively used  (see a recent review in Ref.  \cite{doi:10.1021/cr200148b})
in a wide variety of problems including electron transfer reactions 
\cite{doi:10.1021/acs.jctc.6b01085,doi:10.1021/ct0503163,Segal2007,PhysRevB.93.165102,
doi:10.1021/ct100508y}, excitation energy transfers \cite{Yost2014},
calculation of coupling parameters \cite{PhysRevB.78.165108}, and
non-collinear magnetism \cite{PhysRevB.91.054420}. Computational approaches 
using local constraints \cite{PhysRevB.75.115409}, orthogonality
conditions \cite{doi:10.1021/jp401323d}, and constrained-orbitals 
\cite{doi:10.1021/acs.jctc.7b00362} have also been developed. 

In the direct optimization approach of Wu and Van Voorhis \cite{PhysRevA.72.024502},
a constraint is added to the energy functional using the Lagrange
multiplier method. The Lagrange multiplier determines the constraining 
potential but it is not explicitly known. Wu and Van Voorhis have
shown that the functional is a strictly concave function of the Lagrange
multiplier and there is a unique stationary point which is a maximum. 
They proposed a nested loop approach with an outer self-consistent loop (a normal DFT
loop) and an inner constraint loop. The constraint loop determines the
Lagrange multiplier aided by the first and the second derivatives of the
functional. The constraint iterations are relatively cheap using localized orbitals
(the cost is a diagonalization of the Hamiltonian), but in the case of plane wave or 
real space grid codes describing larger systems, this step can be a bottleneck.

In this paper we implement an approach that simultaneously minimizes
the energy and optimizes  the Lagrange multipliers to satisfy the 
constraints.  The  method uses steepest descent iteration for the
orbitals and the Lagrange multiplier is iteratively updated in each 
step. The Lagrange multiplier is adjusted in each iteration in order
to enforce  the constraining condition \cite{Cusson1985,PhysRevC.32.172}
on the desired expectation value. The advantage of the approach is 
that it can be easily implemented
alongside steepest descent or conjugate gradient minimization
allowing efficient cDFT calculations using real space grids. 
A distinctive merit of the method is that it can also be used to
enforce spatially varying constraints. One can constrain not only a
prescribed total charge in a region, but a desired density
distribution can also enforced opening new possible applications for 
cDFT. 

In section \ref{sec:form}, we outline the main points of the formalism,
leaving the details collected in the Appendices. In section
\ref{sec:res}, numerical tests will be presented. The last section is a
short summary. Two appendices are added to describe   the iterative
diagonalization formalism, to overview of Lagrange multiplier
approach, and to motivate the iterative optimization.


\section{\label{sec:form}Formalism}

\subsection{Constrained Density Functional Theory}

In DFT the total energy in atomic units (a.u.) is given by
\begin{equation}
 E[\rho]=T+\int d\bf{r}\, \textit{v}_{\textit{n}}(\bf{r})\rho(\bf{r})+
 {\textit J}[\rho]+{\textit E}_{\text{xc}}[\rho^{\alpha},\rho^{\beta}],
\label{ks-energy}
\end{equation}
where 
\begin{equation}
T=\sum^{\alpha,\beta}_{\sigma}\sum_i^{N_{\sigma}}\langle{\psi_{i\sigma}}\vert-\frac{1}{2}\nabla^2\vert
{\psi_{i\sigma}}\rangle
\end{equation}
is the kinetic energy,  $J$ is the Coulomb energy, $E_{\text{xc}}$ is the exchange-correlation energy, 
$\textit{v}_n(\bf{r})$  is the external potential, and
\begin{equation}
\rho^\sigma({\bf r})=\sum_i^{N_{\sigma}}\left| \psi_{i\sigma}({\bf r})\right|^2
\end{equation}
is the electronic density for spin 
$\sigma=\uparrow, \downarrow$ of $N_{\sigma}$ electrons 
($\rho=\rho^\uparrow+\rho^\downarrow$).
A generic constraint can be defined as
\begin{equation}
\sum^{\alpha,\beta}_{\sigma}\sum_i^{N_{\sigma}}\langle{\psi_{i\sigma}}
\vert {\hat Q}^\sigma \vert
{\psi_{i\sigma}}\rangle=Q_0,
 \label{constr}
\end{equation}
where ${\hat Q}^\sigma({\bf r})$ is an operator and $Q_0$ is a desired expectation
value. For example, it is very common to constrain the charge density
so that there is a specified number of electrons for each spin, $N_\text{c}^\sigma$, within a certain region 
of space:
\begin{equation}
\int w^{\sigma}_{\text{c}}(\bf{r})\rho^{\sigma}(\bf{r})\textit{d}\bf{r}=\textit{N}_\mathrm{c}^\sigma.
\end{equation}
Here, $w_\mathrm{c}^\sigma(\bf{r})$ is a weighting function confining
the electron density into a specified  spatial  region
(e.g. $w_\mathrm{c}^\sigma(\bf{r})$ is equal to 1 within a certain volume and 0 elsewhere).

In order to minimize the  total  energy of Eq.~\ref{ks-energy} subject to the constraint of Eq.~\ref{constr}, 
a functional is defined to be
\begin{equation}
L[\rho,\lambda]=E[\rho]+\lambda \left(\sum^{\alpha,\beta}_{\sigma}\sum_i^{N_{\sigma}}\langle{\psi_{i\sigma}}
\vert {\hat Q}^\sigma \vert {\psi_{i\sigma}}\rangle-Q_0
\right),
\label{functional}
\end{equation}
where $\lambda$ is the Lagrange multiplier. 

Minimizing $L$ with respect to $\lambda$ forces the constraint to  be
satisfied.  By making this functional stationary under the condition that the
Kohn-Sham orbitals are orthonormalized (see Appendix \ref{con} and the
discussion in Ref. \cite{PhysRevB.94.035159}), one gets the Kohn-Sham equations
with an extra term, the constraining potential $\lambda Q^\sigma({\bf r})$,
\begin{equation}
\left({\hat H}_\text{KS}^\sigma+\lambda Q^\sigma ({\bf r})\right)
\psi_i^\sigma({\bf r})=\epsilon_i\psi_i^\sigma({\bf r}). 
\label{cks}
\end{equation}
Here,
\begin{equation}
{\hat H}_{KS}^\sigma=
-\frac{1}{2}\nabla^2 + \textit{v}_\textit{n}(\bf{r})+\textit{v}_{\text{xc}}^\sigma(\bf{r})+
\int \frac{\rho(\bf{r'})}{\left|\bf{r}-\bf{r'}\right|}\textit{d}\bf{r'},
\end{equation}
where $\textit{v}_{\text{xc}}^\sigma$ is the exchange and correlation
potential. Up until now, the popular notation of the
literature has been followed; however, from now on, we drop the spin index for simplicity and
assume that each orbital is doubly occupied. 

For a given $\lambda$, one can determine the orbitals, and with the correct
$\lambda$ the constraint is fulfilled. 
Wu and Van Voorhis
\cite{PhysRevA.72.024502} have
established a means of solving for a unique stationary point.
They have shown that $L(\rho, \lambda)$ is a strictly concave function 
of $\lambda$, with only one
stationary point which is a maximum. Both the first and second
derivatives of $L$ with respect to $\lambda$ can be derived, so the
optimization can be done efficiently. Finding $\lambda$ requires the
solution of Eq. $\eqref{cks}$ for a given lambda and updating lambda, thus optimizing $L$
(see Appendix \ref{WvV}).
The desired constraining potential is found when the constraining
equation is satisfied with respect to a prescribed accuracy. Further discussion on the optimization of
constrained DFT can be found in Ref. \cite{PhysRevB.94.035159}, in which efficient
calculations involving multiple constraints are described.

\subsection{Iterative minimization}
We will use a method that is based on iterative diagonalization. This
approach is often used in cases of large basis dimension, such as for the three-dimensional real space grid 
representation, where direct diagonalization of the Hamiltonian matrix is infeasible and
alternate methods must be used to determine the lowest energy eigensolutions. 
The simplest approach is a steepest descent iteration
\begin{multline}
\psi_j^{(n+1)}(\rvec) = \mathcal{O}\left\{\psi_j^{(n)}(\rvec) - \right. \\ 
\left. x_0\left({\hat H}_\text{KS}-\epsilon_j^{(n)}\right)\psi_j^{(n)}(\rvec)\right\}.
\end{multline} 
Here, $x_0=\Delta t / \hbar$,
\begin{equation}
\epsilon_j^{(n)}=\langle \psi_j^{(n)}\vert {\hat H}_\text{KS} \vert
\psi_j^{(n)}\rangle,
\end{equation}
and $\mathcal{O}$ indicates Gram--Schmidt orthonormalization, required to preserve the 
orthonormality of the single-particle states at each update step. The
starting wave function, $\psi_j^{(n)}$, is some initial guess, e.g.
linear combination of atomic orbitals, and $x_0$ is chosen to be sufficiently
small for convergence. The steepest descent step can be derived from
imaginary time propagation and can be improved by
using higher order approximations to the exponential operator (see Appendix \ref{itp}).

\subsection{Iterative minimization with a constraint}
The advantage of the iterative diagonalization is that it can be
combined with a step which is designed to enforce the constraints. The motivation for the
concrete form of the iterative updates, the possible implementations,
and the highlights of earlier works is summarized in Appendix \ref{con}. 

In the case of constraint, the goal is to update each orbital towards the minimum energy 
configuration while maintaining that an arbitrary expectation value, related to an 
associated operator, ${\hat Q}$, does not change from one static iteration to the next; i.e.
\begin{equation}
\sum_j \langle \psi_j^{(n+1)} | {\hat Q} | \psi_j^{(n+1)} \rangle =\sum_j \langle \psi_j^{(n)} | {\hat Q} | \psi_j^{(n)} \rangle. 
\label{cond}
\end{equation}
Furthermore, the value of this expectation value is meant to match a given input value,
\begin{equation}
\sum_j\langle \psi_j^{(n+1)} | {\hat Q} | \psi_j^{(n+1)} \rangle = Q_0. 
\label{cond1}
\end{equation}
These conditions may be incorporated into the above iterative formalism by the inclusion of a Lagrange
multiplier constraint term such that the new update scheme becomes
\begin{multline}
\psi_j^{(n+1)}(\rvec) = \mathcal{O}\left\{\psi_j^{(n)}(\rvec) - \right. \\ \left. x_0\left({\hat H}_\text{KS} + \lambda^{(n)}
{\hat Q} - 
\epsilon_j^{(n)}\right)\psi_j^{(n)}(\rvec)\right\}.
\end{multline}
In this new update scheme, one has to simultaneously  iterate the
Lagrange multiplier $\lambda^{(n)}$. The simplest choice is to use a
steepest descent iteration for $\lambda$ as well (see Appendix \ref{io}), but one can work
out a much better scheme by choosing $\lambda^{(n)}$ in such a way that
the the constraint in Eq. \eqref{cond} is satisfied.  

To this end
\cite{Cusson1985,PhysRevC.32.172}, one includes  an intermediate step
\begin{multline}
\psi_j^{(n+1/2)}(\rvec) = \mathcal{O}\left\{\psi_j^{(n)}(\rvec) -
\right. \\ \left. x_0\left({\hat H}_\text{KS} + \lambda^{(n)}{\hat Q} - \epsilon_j^{(n)}\right)\psi_j^{(n)}(\rvec)\right\}.
\label{step1}
\end{multline}
The difference of the relevant expectation value between the original and half steps is calculated,
\begin{equation}
\delta Q = \sum_j\langle \psi_j^{(n+1/2)} | {\hat Q} | \psi_j^{(n+1/2)}
\rangle -\sum_j \langle \psi_j^{(n)} | {\hat Q} | \psi_j^{(n)} \rangle, 
\end{equation}
so that the Lagrange multiplier may be updated as
\begin{eqnarray}
\lambda^{(n+1)} = \lambda^{(n)} &+& 
c_0 \frac{\delta Q}{2x_0\sum_j \langle \psi_j^{(n)} | {\hat Q}^2 |
\psi_j^{(n)} \rangle + d_0}\nonumber\\
&+&\frac{\sum_j\langle \psi_j^{(n)} | {\hat Q} | \psi_j^{(n)} \rangle - Q_0}{2x_0\sum_j \langle \psi_j^{(n)} | {\hat Q}^2 | \psi_j^{(n)} \rangle + d_0}.
\label{step2}
\end{eqnarray}
Here, $c_0$ and $d_0$ are numeric constants; their role is explained
in Appendix \ref{cio}. 
A good choice for $c_0$ is a value between 0.9 and 1.0, and that for $d_0$ is around $7 \times 10^{-5}$. 
In the above equation, $\lambda$ is corrected with two terms. The first correction seeks to preserve 
the expectation value of ${\hat Q}$ by reducing the change in $\delta
Q$  (see Eq. \eqref{blam2}). The second correction term adjusts the expectation value toward
the desired value (see Eq. \eqref{blam1}).

With these readjustments, the $(n+1)$th step is given as
\begin{multline}
\psi_j^{(n+1)}(\rvec) = \mathcal{O}\left\{\psi_j^{(n+1/2)}(\rvec) 
- \right. \\ \left. x_0\left(\lambda^{(n+1)} - \lambda^{(n)} + \delta\lambda\right){\hat Q}\psi_j^{(n+1/2)}(\rvec)\right\},
\label{step3}
\end{multline}
This update step can be considered as a simultaneous correction meant to
preserve the expectation value as well as force the expectation value
to be equal to a desired quantity. The numerical constants appearing in
the iteration play a similar role to the density mixing parameters 
in the self-consistent solution of the Kohn-Sham equations by helping
the speed of convergence. The motivation and details of the above steps for the simultaneous
diagonalization of the Hamiltonian and the optimization of $\lambda$ is given in Appendix
\ref{cio}. 

This update scheme settles the Kohn--Sham system into the minimum energy state 
while maintaining a constraint on an arbitrary state expectation value. Effectively, 
what occurs is the convergence of the Kohn--Sham system towards the global ground state 
for a total effective potential which is iteratively updated simultaneous to the orbitals. 
Thus, the final state may be fully constructed by real-valued orbitals, and the converged 
Lagrange multiplier term, $\lambda^{\rm final}Q$, represents a fictitious, additional 
external potential which corresponds to a Kohn--Sham state exhibiting the desired 
expectation value.


\section{\label{sec:res}Results}
In this section we present results of the iterative  constraint 
update scheme. In each case, a real space 
grid representation was used alongside a finite difference representation of the kinetic energy. 
The ion cores were treated using norm-conserving Troullier and Martins pseudopotentials 
\cite{PhysRevB.43.1993}.

\subsection{Simple model system}
\begin{figure}[ht]
	\includegraphics[width=.45\textwidth]{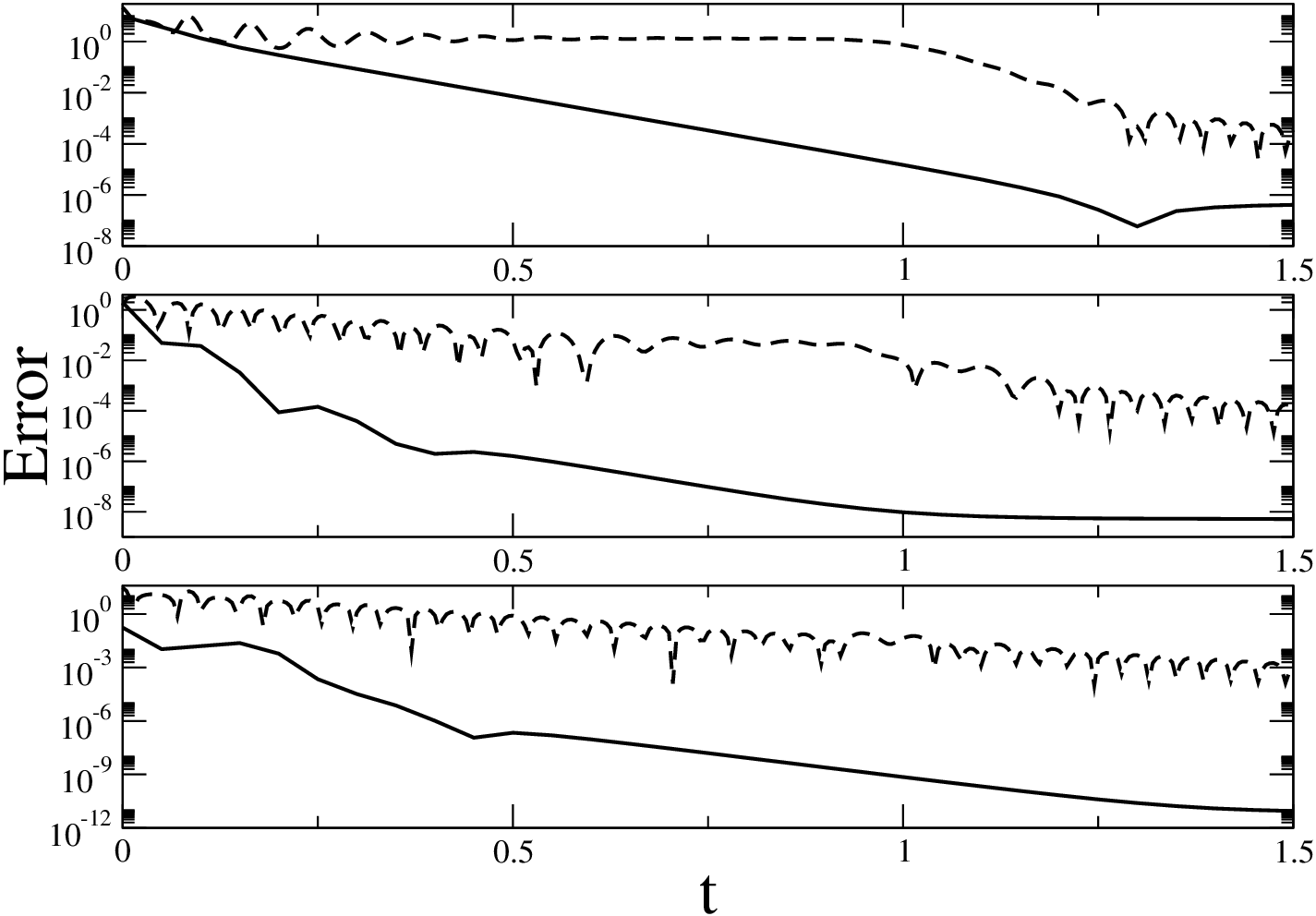}
	\caption{Convergence of energy (top),  Lagrange multiplier,
    $\lambda$, (middle),
    and $Q$ (bottom) as a function of  time (in a.u.)  for a
    simple steepest decent update (dashed line, $x_0=0.0005$ a.u.) and the $\lambda$
    optimization (solid line, $x_0=0.005$ a.u.) }
\label{fig:ho}
\end{figure}

As a simple numerical test we consider a three dimensional harmonic
oscillator $V({\bf r})={1\over 2} \omega^2{\bf r}^2$ (a.u.) 
with $N$=5 orbitals, subject to the constraint
\begin{equation}
Q-Q_0=0, 
\end{equation}
where 
\begin{equation}
Q=\sum_{j=1}^N \langle \psi_j^{(n)} | {\bf r}^2 | \psi_j^{(n)} \rangle. 
\end{equation}
In the test calculation, the parameters $\omega$ and $Q_0$ are chosen to be 1 and 25 a.u., respectively. 
The model is analytically solvable. Adding
the $\lambda Q$ term to the Hamiltonian is equivalent to a modified
harmonic oscillator potential with $\omega'=\sqrt{\omega^2+\lambda}$.
The square radius of a harmonic oscillator wave function with quantum
numbers $(n_x,n_y,n_z)$ is equal to ${1\over 2\omega}(2n_x+2n_y+2n_z+3)$,
so the condition $Q_0=25$ determines the analytical value of $\lambda$
and the energy. The numerical solution for a simple steepest descent
update and the $\lambda$ optimization approach presented
in the previous section is compared in Fig. \ref{fig:ho}. The figure
shows that the $\lambda$ optimization is very accurate, both in energy
and in constraining $Q$, and  the $\lambda$ convergence is very fast. \revDK{The steepest
descent approach, based on Eqs. \eqref{gradstep} and \eqref{las},  also works but the accuracy is orders of
magnitudes worse.} This simple but clean example (no self consistency)
shows that the $\lambda$ optimization approach is accurate and fast.

\subsection{Charge constraint}
Now we apply the iterative optimization scheme to charge  transfer systems 
studied by  Wu and Van Voorhis  \cite{PhysRevA.72.024502, doi:10.1021/ct0503163}. 
They used a weight function,
$w(\rvec)$, which designates coordinate space belonging to the donor with a value 
of 1 and that of the acceptor with a value of -1. In this way, a Lagrange multiplier term is added 
to the Kohn--Sham equation as $\lambda Q=\lambda w(\rvec)$, effectively representing a 
step potential which may be tuned during optimization until the desired 
charge imbalance between the two partitions,
\begin{equation}
N_c =Q_0 =
\sum_{j=1}^N \langle \psi_j^{(n)} | w({\bf r}) | \psi_j^{(n)} \rangle =
\int w(\rvec)\rho(\rvec)d\rvec,
\end{equation}
is reached. 
The weight function may be defined using a scheme such as Hirshfeld partitioning \cite{doi:10.1063/1.3507878, Hirshfeld1977} such that 
\begin{equation}
w(\rvec) = \frac{\sum_{i\in D}\rho_i(\rvec-\Rvec_i) - \sum_{i\in A}\rho_i(\rvec-\Rvec_i)}{\sum_{i}\rho_i(\rvec-\Rvec_i)},
\end{equation}
where $\rho_i(\rvec)$ represents the unperturbed electron density of ion $i$ and $\Rvec_i$ is its location.

In our formalism, the weight function may be chosen as the operator whose associated expectation value, $N_c$, is being constrained to a given value. In this way, charge constraint optimization may be performed in DFT using a real space grid approach which, unlike the atomic orbitals basis, does not allow for a practical means of storing the full Hamiltonian matrix and, instead, relies on algorithms which describe the action of the Hamiltonian matrix on a wave function vector. Furthermore, in this update scheme, one is not required to use a nested-loop form in which either the energy minimization or the constraint condition is satisfied via an inner loop while the other is satisfied using the outer loop. 
In the iterative constraint method, one progresses towards the stationary point by simultaneously updating each. 
This may lead to significantly faster runtimes or enhanced stability.

\begin{figure}[ht]
	\includegraphics[width=.45\textwidth]{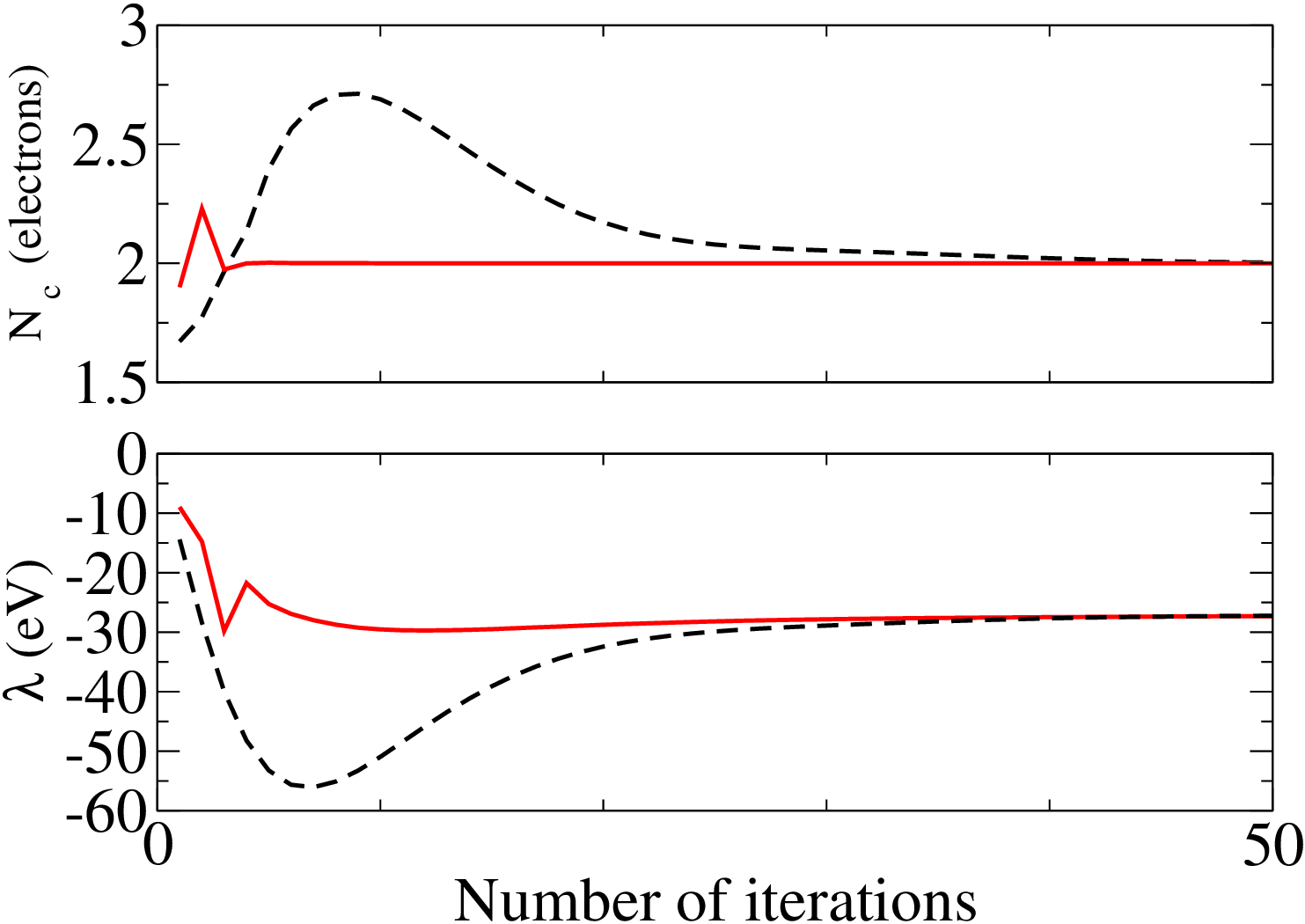}
	\caption{Convergence of charge difference (top) and Lagrange
    multiplier (bottom) for a ${\rm N}_2$ molecule. The input desired charge difference was $N_c = 2$ electrons.
    The black dashed line is obtained using the approach of Ref.
    \cite{PhysRevA.72.024502}, the red solid line is the result of the present
    approach.
    }
\label{fig:N2_con}
\end{figure}

\begin{figure}[ht]
  (a) \hspace{.15\textwidth} (b) \hspace{.15\textwidth} (c) \\
	\includegraphics[width=.15\textwidth, trim={4cm 9cm 4cm 9cm},clip]{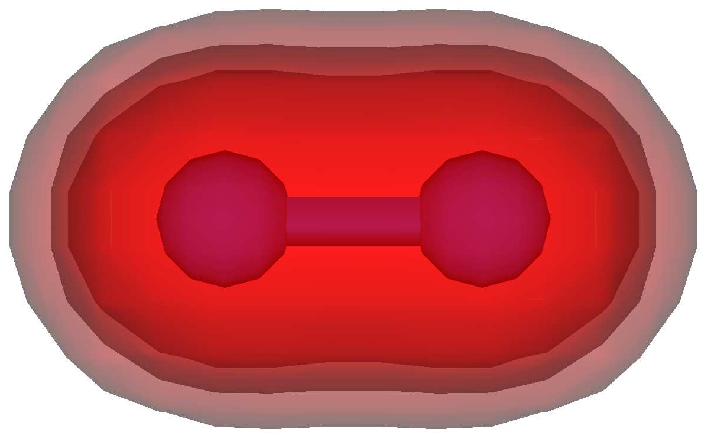}
	\includegraphics[width=.15\textwidth, trim={4cm 9cm 4cm 9cm},clip]{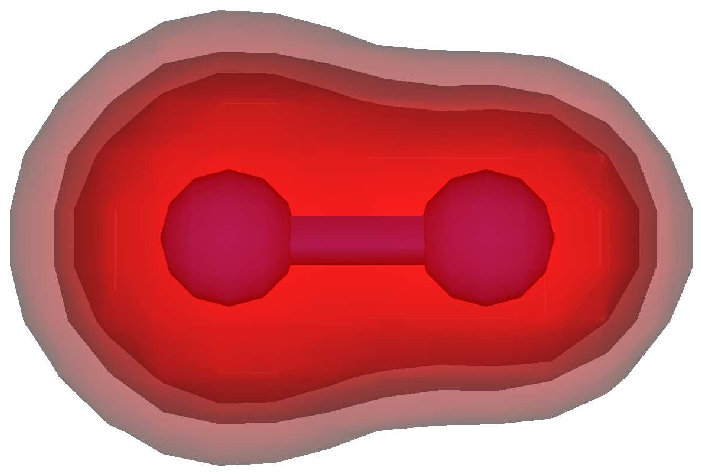}
	\includegraphics[width=.15\textwidth, trim={4cm 9cm 4cm 9cm},clip]{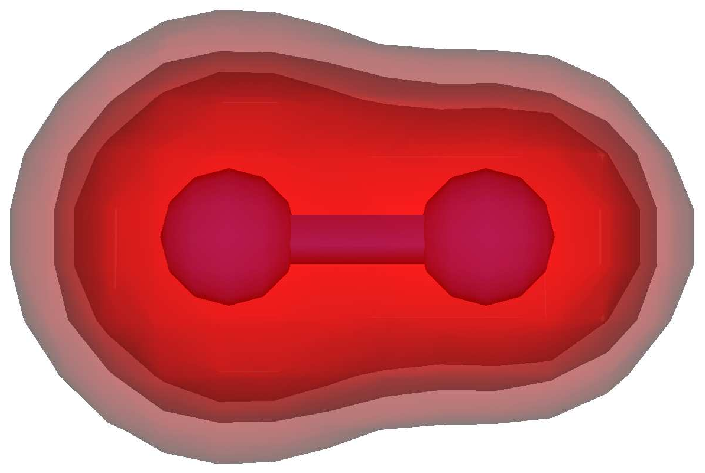}
	\caption{Converged electron density of a ${\rm N}_2$ molecule
    calculated using (a) conventional DFT, (b) constrained DFT with an
    imposed charge difference of 2 electrons, and (c) conventional DFT
    with an additional external potential of 
    $-27.01 {\rm eV} \times w(\rvec)$. Three isosurfaces corresponding to the density of 
    0.33, 0.67, and 1.0 ${\rm \AA}^{-3}$ are shown.}
\label{fig:N2_den}
\end{figure}

Of the simplest cases to consider is the diatomic ${\rm N}_2$
molecule. Here, one atom is designated as the donor, and the other is
the acceptor. The above described procedure was carried out for a
desired charge difference between the two atoms of $N_c \rightarrow 2$
electrons. The width of the computational box was 6 Angstroms on each side with 25 grid points along each axis. A plot of the convergence of $N_c$ and of the Lagrange multiplier, $\lambda$, is presented in Fig. \ref{fig:N2_con}. We find that only a small number of iterations are needed for satisfactory convergence in this case. The resulting value for $\lambda$, indicating the depth of the step potential enforcing the charge difference, was -27.01 eV. 
The electron density for the ${\rm N}_2$ molecule using conventional
DFT is shown in Fig. \ref{fig:N2_den}(a) and that of the charge
constrained ${\rm N}_2$ molecule is shown in Fig. \ref{fig:N2_den}(b).
By including an additional potential of $-27.01 {\rm eV} \times w(\rvec)$ in a conventional DFT calculation of the ${\rm N}_2$ molecule, the charge difference of $N_c=2$ electrons naturally arises and the density, shown in Fig \ref{fig:N2_den}(c), nearly exactly matches that of the constrained DFT case. 

The present approach and that of Wu and Van Voorhis is  compared in Fig.
\ref{fig:N2_con}. The present approach converges much faster for
$N_c$, after about 20 iterations the value of $N_c$ is accurate up to
5-6 digits. More importantly,
$\lambda$ also converges faster using the optimized iteration, despite
the fact that it's value is adjusted to enforce the constraint. To
compare the computational burden, we note that each self consistent
loop was updated in the same way for both approaches. The computational
cost difference comes from the fact that the Wu and Van Voorhis method
is using an internal loop (see Appendix \ref{WvV}) \rev{which requires  3-4 additional
$\left({\hat H}_\text{KS} + \lambda^{(n)}{\hat Q}\right)\psi_j^{(n)}(\rvec)$
operations per iteration than in our implementation.} 
Thus the computational cost of the Wu and Van
Voorhis approach is about 3-4 times higher than that of the present one for the 
same number of self-consistent energy minimization iterations. 

This comparison is meant to highlight the potential efficiency of the
present method, but it is not a strict comparison of the 
computational cost. Using the two approaches for different systems or
using different basis functions might result in different computational 
efficiencies. Calculations using localized basis functions with small 
Hamiltonian matrices would most definitely be faster using the method of Wu and Van Voorhis.
The second derivate (see  Appendix \ref{WvV}) of the functional $L$
would also increase the convergence, but that is not readily available
in real space calculations.

We next consider the small systems tested by Wu and Van Voorhis in Table 1
of Ref. \onlinecite{doi:10.1021/ct0503163}. For long separation distances between the donor and acceptor molecules, $\Rvec$, one would expect that the energy varies as $1/\Rvec$. A good test of the energies calculated by a charge constraint DFT program would be to plot the total energy vs. $1/\Rvec$ and show the expected linear dependence. Furthermore, one may use the slope of this curve in order to determine the total energy of the charge-separated (CS) system; i.e. $\Rvec \rightarrow \infty$. Being that at 
an infinite separation distance, the two molecules should be independent of one another, this total energy should equal the sum of separate calculations for the appropriately ionized donor and acceptor molecules, that is ${\rm D}^-$ and ${\rm A}^+$ in the case of $N_c=2$. 

\begin{figure}[ht]
	\includegraphics[width=.45\textwidth]{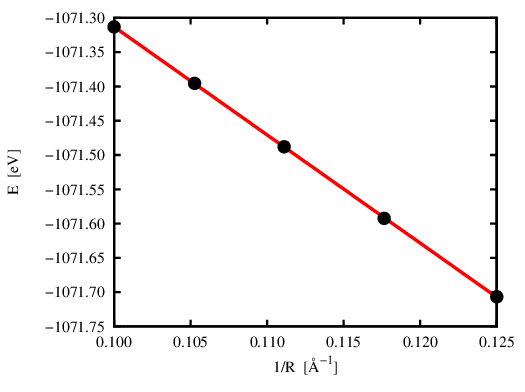}
	\caption{Total energy calculations using charge difference constrained DFT to describe the system ${\rm N}_2^+$--${\rm N}_2^-$ for five separation distances, $\Rvec$, chosen within a range of 8 to 10 {\AA}.}
\label{fig:N2-N2}
\end{figure}

An example trend of the total energy vs. $1/\Rvec$ is presented in Fig. \ref{fig:N2-N2} for the system ${\rm N}_2^-$--${\rm N}_2^+$ and a charge difference of $N_c=2$. The data is well fit by a linear trend line. The slope indicates an expected CS total energy of -1069.735 eV. This value is in good agreement with the total energy of separate calculations for the ${\rm N}_2^+$ and ${\rm N}_2^-$ molecules, being -1070.970 eV. These values, as well as those for the cases of ${\rm H}_2{\rm O}^-$--${\rm F}_2^+$ and 
${\rm C}_2{\rm F}_4^-$--${\rm C}_2{\rm H}_4^+$, are presented 
in Table \ref{tab:small_mol}. 
We note that each value for $\left( E_{{\rm D}^-} + E_{{\rm A}^+}\right)$ is higher 
than its $E_{\rm CS}$ counterpart by about 1.15 eV. Apart from this small systematic 
shift, all values agree well, indicating that these long-range charge transfer states are being well-represented by the present scheme. 
\revDK{We note that the reported values in Table \ref{tab:small_mol} do not represent total energies of these systems. This is due to the fact that pseudopotentials have been employed, and the frozen-core approximation energies, corresponding to the pseudopotential contributions, have been neglected. In principle, because the difference in total energy is nearly always the desired calculated quantity, one may compare such valence energies among like systems, represented using the same pseudopotentials, as if they were the true total energies.}    

\begin{table}[ht]
\begin{tabular}{cc|ccc}
 D                      & A                      & $E_{\rm CS}$ & $\left( E_{{\rm D}^-} + E_{{\rm A}^+}\right)$ & \% diff. \\\hline
 ${\rm N}_2^-$          & ${\rm N}_2^+$          & -1069.735    & -1070.970                                     &  0.115   \\
 ${\rm H}_2{\rm O}^-$   & ${\rm F}_2^+$          & -1767.172    & -1768.204                                     &  0.058   \\
 ${\rm C}_2{\rm F}_4^-$ & ${\rm C}_2{\rm H}_4^+$ & -3323.066    & -3324.250                                     &  0.036   
\end{tabular}
\caption{The charge separated state energy, $E_{\rm CS}$, for three small molecule cases, determined using a linear fit of five data points representing the total energy found using a charge difference constraint of $N_c=2$ for separation distances ranging between 8 and 10 Angstroms. These values are compared to the sum of individual calculations for each constituent ionized molecule.}
\label{tab:small_mol}
\end{table}

Finally, we investigate the popular charge transfer excitation in zincbacteriochlorin-bacteriochlorin (ZnBC--BC). This system is a common component of suggested light harvesting devices, appearing with a phenylene link. However, it has been demonstrated that ignoring the link introduces negligible error \cite{doi:10.1021/ja039556n}; therefore, the pair of isolated molecules separated by 5.84 {\AA} is commonly studied.  The charge transfer excited state of this complex was of the earliest shown to be misrepresented by
time-dependent density functional theory (TDDFT)
\cite{doi:10.1021/ja039556n}, due to an incorrect treatment of the long-range exchange potential. Thus, there are many studies devoted to the correction of this shortcoming. Such calculations include TDDFT using various local \cite{doi:10.1021/ja039556n, doi:10.1063/1.1461815, KOBAYASHI2006106} and hybrid \cite{KOBAYASHI2006106} functionals, methods combining a configuration interaction singles (CIS) approach \cite{doi:10.1021/ja039556n, doi:10.1021/jp068409j}, and also use of the Bethe--Salpeter formalism \cite{PhysRevLett.109.167801}. The CDFT formalism of Wu and Voorhis was also initially applied to this system \cite{PhysRevA.72.024502, doi:10.1021/ct0503163}, there using an atomic orbitals basis, specifically the ${\rm 6-31G}^*$ basis set, and the 
Becke--Lee--Yang--Par (BLYP) functional
\cite{PhysRevA.38.3098,PhysRevB.37.785}. This approach has also been recently tested on the ZnBC-BC complex using a flexible Daubechies wavelet basis and the LDA functional \cite{doi:10.1063/1.4922378}.

\begin{figure}[ht]
\begin{minipage}{0.1\linewidth}
(a)
\end{minipage}\hfil
\begin{minipage}{0.8\linewidth}
    \includegraphics[width=\linewidth, trim={2cm 9cm 1cm 10cm},clip]{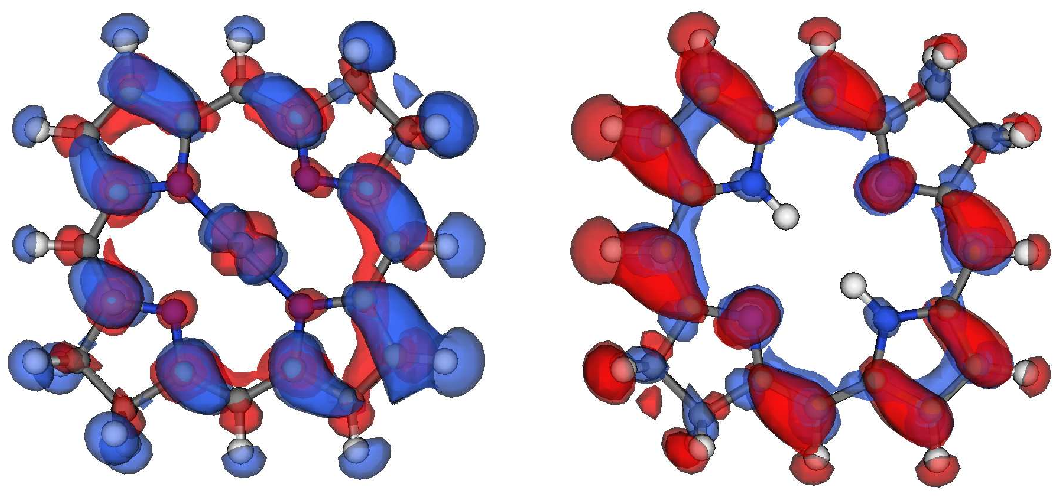}
\end{minipage}
\begin{minipage}{0.1\linewidth}
(b)
\end{minipage}\hfil
\begin{minipage}{0.8\linewidth}
    \includegraphics[width=\linewidth, trim={2cm 9cm 1cm 10cm},clip]{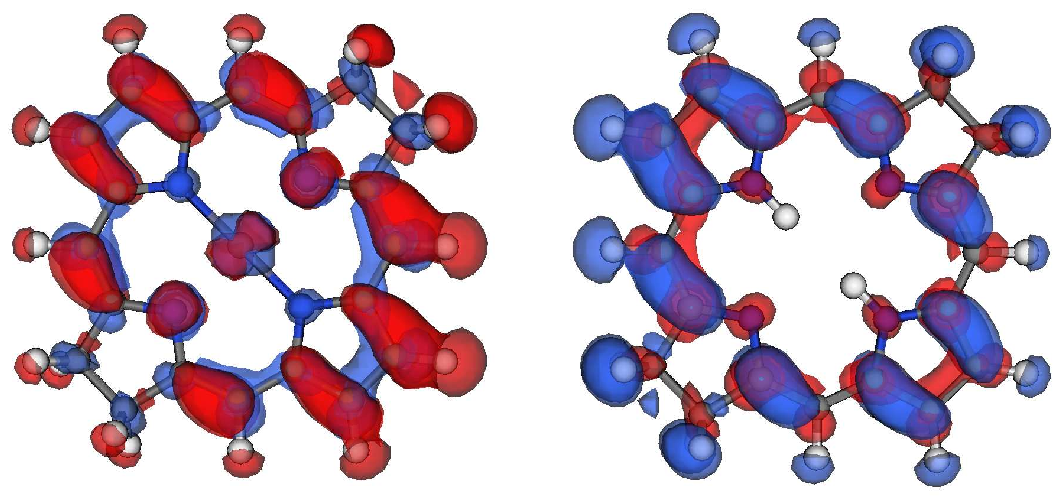}
\end{minipage}
	\caption{Density difference between a ground state DFT calculation of ZnBC--BC and charge constrained DFT representing (a) ${\rm ZnBC}^+$--${\rm BC}^-$ and (b) ${\rm ZnBC}^-$--${\rm BC}^+$. One isosurface corresponding to densities of 0.006 ${\rm \AA}^{-3}$ are shown. Red (blue) indicates positive (negative) values.}
\label{fig:ZnBC-BC}
\end{figure} 

\begin{figure}[ht]
	\includegraphics[width=.45\textwidth]{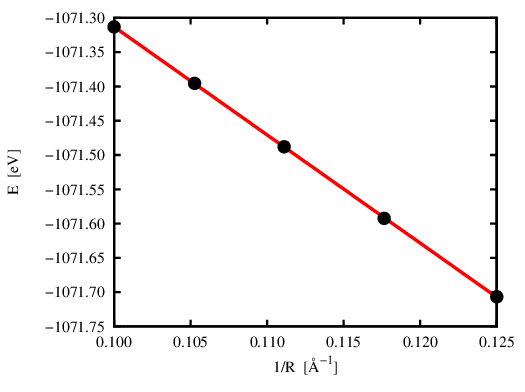}
	\caption{Total energy calculations using charge difference constrained DFT to describe the systems ${\rm ZnBC}^+$--${\rm BC}^-$ and ${\rm ZnBC}^-$--${\rm BC}^+$ for seven separation distances, $\Rvec$, chosen at equal increments within a range of 5.84 to 8.84 {\AA}.}
\label{fig:ZnBC-BC_lf}
\end{figure} 

The difference of the DFT-calculated ground state density for ZnBC--BC and the charge constrained DFT density is shown in Fig. \ref{fig:ZnBC-BC}(a) and Fig. \ref{fig:ZnBC-BC}(b) for the cases of ${\rm ZnBC}^+$--${\rm BC}^-$ and ${\rm ZnBC}^-$--${\rm BC}^+$, respectively. 
The LDA relaxed coordinates were provided by Ratcliff {\etal} of Ref. \onlinecite{doi:10.1063/1.4922378}. The energy values for this large system maintain a linear dependence with regards to $1/\Rvec$, as shown in Fig. \ref{fig:ZnBC-BC_lf}. The energies of the two excited states relative to the neutral ground state, using a separation distance of $\Rvec = 5.84$ {\AA}, were determined to be 3.54 eV for 
${\rm ZnBC}^+$--${\rm BC}^-$ and 3.95 eV for ${\rm ZnBC}^-$--${\rm BC}^+$. 
\rev{The latter value agrees well with previously calculated values 
(3.91 eV \cite{doi:10.1021/ja039556n}, 3.94 eV
\cite{PhysRevA.72.024502}, and 3.98 eV \cite{doi:10.1063/1.4922378}) 
while the former is lower than similar studies 
(3.71 eV \cite{doi:10.1021/ja039556n}, 3.75 eV
\cite{doi:10.1063/1.4922378}, and 3.79 eV \cite{PhysRevA.72.024502}). 
The difference between the results comes from several sources. The
present code uses pseudopotentials and LDA with real space grid
representation, while the
calculations in Refs. \cite{PhysRevA.72.024502,doi:10.1021/ja039556n} 
are based on all-electron codes with the BLYP functional. The computation in 
Ref. \cite{doi:10.1063/1.4922378} is also based on LDA but uses
Daubechies wavelets, significantly reducing the effect of the coarseness of
the real space grid (e.g. eggbox effect). The most important source of
the difference is that in our real space grid approach the weight
function is not represented in the same way as in the other approaches
using the Voronoi grid \cite{doi:10.1002/jcc.10351} or wavelets. }

\subsection{Density constraint}
The approach can be extended to more general constraints as well. In
this section we demonstrate the ability of the present approach to constrain the
spatial density, requiring that $\rho({\bf r})$ is equal to a
given value, $Q_0({\bf r})=\rho_0({\bf r})$. In this case, the operator ${\hat Q}$
becomes the density operator such that
\begin{equation}
\langle \psi_j^{(n)} | {\hat Q} | \psi_j^{(n)} \rangle= 
\left | \psi_j^{(n)}({\bf r}) \right|^2
\end{equation}
and 
\begin{equation}
\lambda {\hat Q} \psi_j^{(n)} = \lambda({\bf r}) \psi_j^{(n)}({\bf r}).
\end{equation}
Given a desired initial density distribution, $\rho_0({\bf r})$, and using
the steps defined  in Eqs. \eqref{step1}, \eqref{step2}, and \eqref{step3},
one looks for the potential, $\lambda({\bf r})$, which generates 
the Kohn-Sham orbitals, $\psi_j^{(n)}$, so that
\begin{equation}
\sum_j \left| \psi_j^{(n)}({\bf r})\right|^2 = \rho_0({\bf r}).
\end{equation}

\begin{figure}[ht]
\includegraphics[width=.45\textwidth]{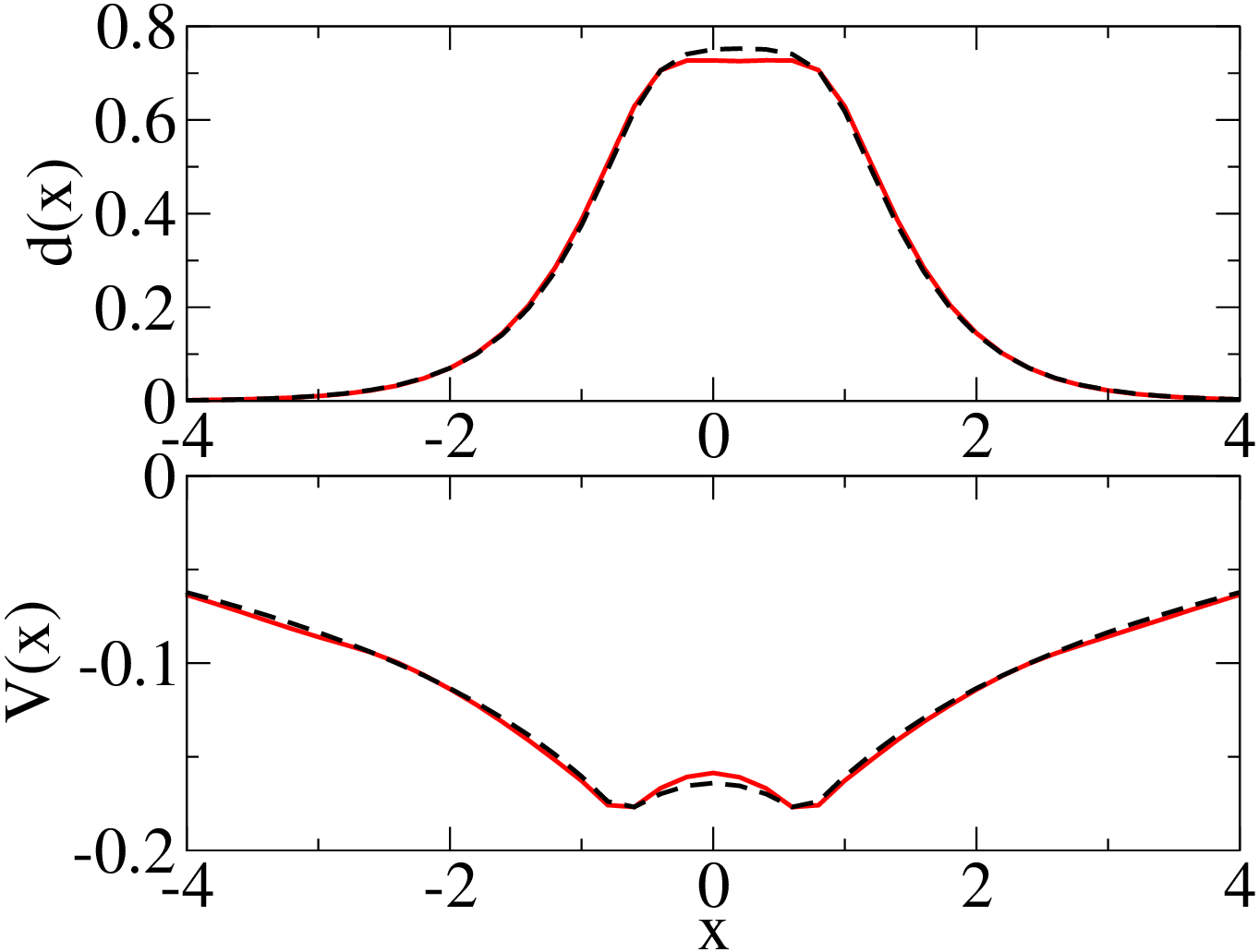}
	\caption{Top: Exact (solid red line) and DFT (dashed black
    line) average density , $d(x)$ along the $x$ axis connecting the
    protons. Bottom: cDFT (solid red line) and DFT (dashed black
        line) average potential, $V(x)$, along the $x$ axis. 
    }
\label{fig:h2}
\end{figure}

As a first example,  we use a simple system, the H$_2$ molecule.
Fixing the protons at 0.74 \AA \ apart, the two-electron Coulomb problem can 
be solved very accurately using the variational method with
explicitly correlated Gaussian basis functions
\cite{RevModPhys.85.693}. The calculated ``exact'' electron density,
shown in Fig. \ref{fig:h2}, will  be the target density $\rho_0({\bf r})$.
Fig. \ref{fig:h2} compares $\rho_0$ to the density obtained by a
conventional DFT calculation. The two densities differ mostly in the
middle region between the two protons where the DFT density is higher. 
Using the density constraint, we then instruct the DFT density to be equal to
$\rho_0({\bf r})$. The asymptotic fall of the density is also different, 
but that is not so important for this test case. 
\revDK{The cDFT calculation constrains the density to satisfy 
${\rm max}|\rho_0({\bf r}) - \rho({\bf r})|<10^{-5}$, and the constrained and exact densities are
indistinguishable in Fig. \ref{fig:h2}.} 
The cDFT potential (the Kohn-Sham potential plus
$\lambda({\bf r})$) and the DFT potential are compared in Fig. \ref{fig:h2}.
The main difference is that the cDFT potential is higher in the middle
region, pushing out the charge and correcting the difference between
the exact and DFT result. In principle, calculations like this can be
used to improve exchange-correlation potentials if accurate
densities are available. To check the calculation, one can use the
resulting $\lambda({\bf r})$ and add it to the Kohn-Sham Hamiltonian
as an external potential. The self-consistent solution produces the
desired density distribution $\rho_0({\bf r})$. 

\begin{figure}[htb]
\centering
  \subfloat[Graphene supercell]{%
    \includegraphics[trim=180 100 180 100, clip, width=.22\textwidth]{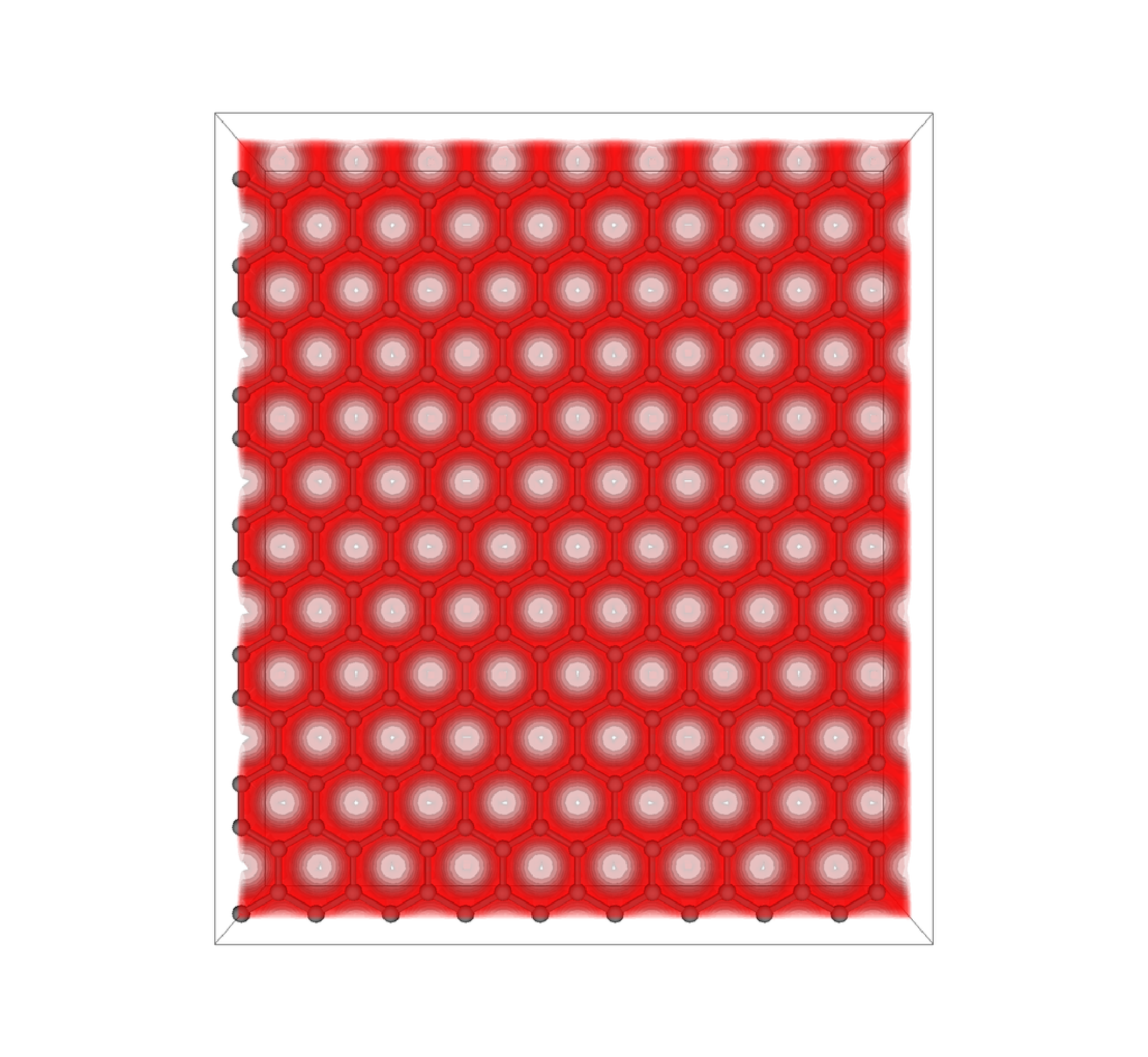}}\hfill
  \subfloat[Target density distribution $\rho_0(\rvec)$]{%
    \includegraphics[trim=180 100 180 100, clip, width=.22\textwidth]{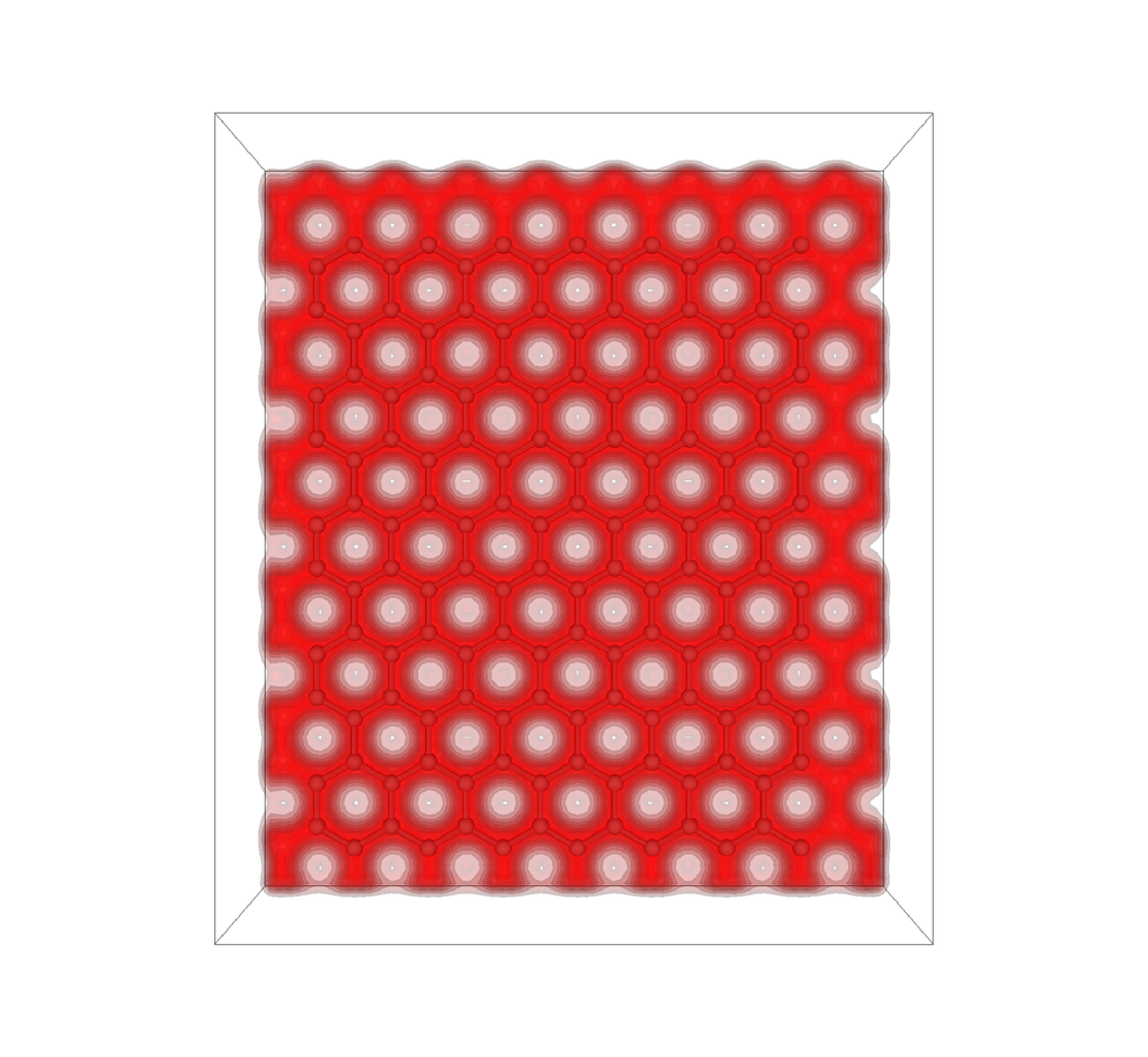}}\hfill
  \subfloat[Generated $\lambda(\rvec)$]{%
    \includegraphics[trim=180 100 180 100, clip, width=.22\textwidth]{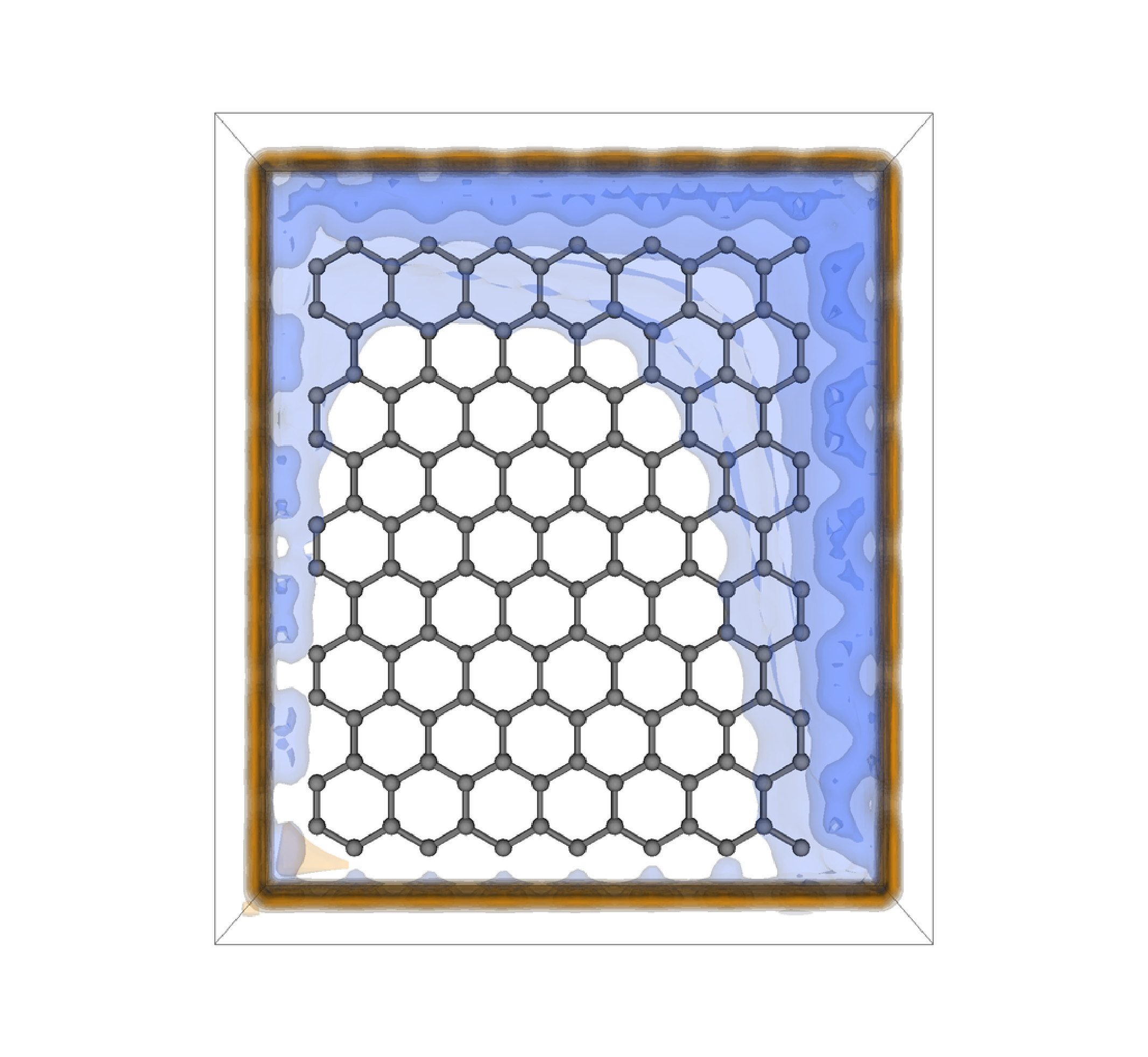}}\hfill
  \caption{Graphene supercell, target density distribution, and $\lambda(\rvec)$ resulting from cDFT. 
           The target density distribution is shown using the graphene fragment molecular geometry used 
           during the cDFT calculation. In depicting $\lambda(\rvec)$, positive (negative) values are
           represented using orange (blue) isosurfaces.}\label{fig:gr}
\end{figure}

The next example demonstrates that the approach works for larger
systems as well. In this case, we calculate the electron density of a
graphene sheet in a periodic supercell calculation. Taking this
density, Fig. \ref{fig:gr}(a), we use a masking function to gradually decrease the
density to zero at the boundaries, Fig. \ref{fig:gr}(b),
and use this distribution as $\rho_0$. 

The cDFT is now used in conjunction with a system of the same molecular 
geometry as the supercell but with the outer perimeter of carbon atoms removed. 
Extra Kohn--Sham orbitals are added, beyond those corresponding to the to 
carbon atoms, in order to ensure that the number of electrons of the initial 
non-converged density matches that of $\rho_0$. We note that there will be one 
orbital of non-integer occupation. The cDFT 
generates $\lambda({\bf r})$, Fig. \ref{fig:gr}(c), so that $H_{\text KS}+\lambda(\rvec)$ 
yields $\rho_0({\bf r})$ as the ground state density. The calculation of a
converged constraining potential needs about two to three times more
iterations than a conventional DFT iteration. \revDK{The calculated
$\lambda({\bf r})$ can be checked by using it as an external potential to
produce $\rho_0({\bf r})$.} 

This example serves to show that the approach is applicable and
converges for larger systems as well. 
One can recognize the formation of potential wells within the shape 
of $\lambda(\rvec)$ near the perimeter of the graphene fragment which 
correspond to the carbon atoms missing in the input molecular geometry. 
In principle this graphene 
fragment can be used to study defects without the problem of
periodic images, but still keeping the proper density. One can also
use the approach to embed a smaller system into a
larger system with density constrains at the boundary. 
In Ref.  \cite{PhysRevB.87.054113} charge densities in a boundary region
between the two domains has been connected using cDFT to facilitate
multiscale calculations using a single scalar $\lambda$ Lagrange
multiplier. The present approach offers a more flexible embedding
possibility.

\section{\label{sec:con}Conclusion}
We have implemented  an iterative optimization approach for constrained
density functional calculations. In this approach the energy
minimization and the optimization of the Lagrange multipliers
enforcing the constraint is simultaneously iterated. The ideal Lagrange
multipliers are determined by enforcing the constraint on the
Kohn-Sham orbitals at each self consistent iteration steps. 

The accuracy and efficiency of the present approach is demonstrated on
previously studied systems. Comparing the computational cost
to previous methods based
on the direct optimization of the Lagrange multiplier 
\cite{PhysRevA.72.024502},  the present approach is expected to be
competitive.  In the case of real space grid basis approaches,
the present method is definitely favorable, by a 2-3 times savings in
computational cost. 

The method is not limited to charge constraints where only a
single (or a small set of) Lagrange multiplier is optimized. We have
shown that one can prescribe a general, spatially varying  density, and 
the external potential that generates this density as a ground state
can be calculated. The applicability of this approach to calculate
a spatially dependent $\lambda({\bf r})$ Lagrange multiplier may open up
possibilities of embedding smaller systems into larger systems, prescribing
boundary conditions using density, or enforcing orthogonality to a given
ground state. 

{\bf Acknowledgement:} The work of A. S. Umar was supported by DOE
grant No. DE-SC0013847.  


\appendix
\section{\label{itp} Imaginary time propagation}
In this appendix, for completeness of the presentation, we show how the
popular steepest descent iteration can be derived from the imaginary time step 
propagation \cite{DAVIES1980111}. \revDK{The important part of this step is that
although we only used, to lowest order, the simplest iteration, systematic 
improvement is possible by including higher order terms.} The origin of 
the imaginary time propagation name comes from the similarity of each optimization 
step and the solution to the time-dependent Kohn--Sham 
equation,
\begin{equation}
i\hbar\frac{\partial}{\partial t}\psi_k(\rvec,t) = {\hat H}_\text{KS}\psi_k(\rvec,t),
\end{equation}
for short time steps, $\Delta t$:
\begin{equation}
\psi_k(\rvec,t+\Delta t) = \text{exp}\left[-i {\hat H}_\text{KS} \Delta t / \hbar\right]\psi_k(\rvec,t).
\end{equation}

For imaginary time step propagation, one makes the transformation $\Delta t \rightarrow -i\Delta t$
and introduces the parameter $x_0=\Delta t / \hbar$ such that the procedure takes the form
\begin{equation}
\psi_j^{(n+1)}(\rvec) = \text{exp}\left[-x_0{\hat H}_\text{KS}\right]\psi_j^{(n)}(\rvec).
\label{eq:imag_time}
\end{equation}
Here, we have replaced the notation of our employed wave functions so that they now 
represent arbitrary functions which are iteratively being updated and approaching the ground
state eigenfunctions of $\psi_k$.
This is seen by noting that such an arbitrary function at iteration $n$, $\psi_k^{(n)}$, may be 
expanded as a linear combination of the eigenfunctions of ${\hat H}_\text{KS}$:
\begin{equation}
\psi_j^{(n)} = \sum_k c_{j,k}^{(n)} \phi_k.
\end{equation}
By plugging this expansion into Eq. \eqref{eq:imag_time}, one obtains 
\begin{equation}
\psi_j^{(n+1)}(\rvec) = \sum_k c_{j,k}^{(n)} \text{exp}\left[-x_0\epsilon_j^{(n)}\right]\phi_k(\rvec)
\end{equation} 
and notes that repeated action by the exponential factor will effectively screen out high 
energy contributions. Thus, if orthonormalization is enforced after each iteration via the
Gram--Schmidt procedure, the functions $\psi_j^{(n)}$ will converge to the ground state
Kohn--Sham orbitals. Equation \eqref{eq:imag_time} is often further 
modified by extracting an arbitrary phase factor from each wave functions which is related to 
their associated Hamiltonian eigenvalue,
\begin{equation}
\psi_j^{(n+1)}(\rvec) = \text{exp}\left[-x_0\left({\hat H}_\text{KS}-\epsilon_j^{(n)}\right)\right]\psi_j^{(n)}(\rvec).
\end{equation}  

In practice, one may approximate the exponential by it's first-order Taylor expansion,
\begin{multline}
\psi_j^{(n+1)}(\rvec) = \mathcal{O}\left\{\psi_j^{(n)}(\rvec) - \right. \\ \left. x_0\left({\hat H}_\text{KS}-\epsilon_j^{(n)}\right)\psi_j^{(n)}(\rvec)\right\}.
\end{multline} 
Here, $\mathcal{O}$ indicates Gram--Schmidt orthonormalization, required to preserve the 
orthonormality of the single-particle states at each update step. 

We note that in order to 
carry out this procedure, one only requires the action of ${\hat H}_\text{KS}$ upon a wave function,
as opposed to needing to store a large matrix. In practice, the damping constant $x_0$ may be 
replaced with a generalized damping matrix, $D(E_0)$ \cite{PhysRevA.40.4182}. Several choices of this operator have 
been investigated \cite{reinhard1982comparative}. That used in this work is of the form
\begin{equation}
D(E_0) = \left[ 1 + \frac{T}{E_0} \right]^{-1},
\end{equation}
where $T$ is the kinetic energy matrix and $E_0$ is a numeric constant. 
A good choice for the latter is the depth of the effective Kohn-Sham potential. 
In this work, determining the action of the damping matrix at each update
step, a problem of the form $\vec{y} = D\vec{x}$, is approximately solved by applying a 
small number of conjugate gradient steps to the equation $\left[ 1 + \frac{T}{E_0} \right]\vec{y} = \vec{x}$.   

\section{\label{con} Constrained system}
\subsection{Lagrange multiplier approach}
We restrict the discussion for a single orbital---the extension for
many orbitals by requiring orthogonality is simple. 
We assume that the wave function is expanded in terms of basis
functions
\begin{equation}
\psi({\bf r})=\sum_{j=1}^K c_{j} \phi_j({\bf r}).
\end{equation}
If if one uses real space grid the basis, 
\begin{equation}
\phi_i({\bf r}_k)=\delta_{ik},
\end{equation}
where ${\bf r}_k$ is a gridpoint, then 
\begin{equation}
\psi({\bf r}_k)=c_{k}.
\end{equation}
We define the matrix elements of the Hamiltonian,
\begin{equation}
H_{ij}= \langle \phi_i \vert {\hat H} \vert \phi_j\rangle,
\end{equation}
overlap,
\begin{equation}
O_{ij}=\langle \phi_i  \vert \phi_j\rangle,
\end{equation}
and constraining operator,
\begin{equation}
Q_{ij}=\langle \phi_i \vert {\hat Q} \vert \phi_j\rangle.
\end{equation}
Using these matrix elements, the energy is
\begin{equation}
E=\sum_{i,j=1}^K c_i c_j H_{ij},
\end{equation}
the norm of the wave function is
\begin{equation}
O=\sum_{i,j}^K c_i c_j O_{ij},
\end{equation}
and the constraint is
\begin{equation}
Q=\sum_{i,j=1}^K c_i c_j Q_{ij}.
\end{equation}
One can now define the functional
\begin{equation}
L(c,\lambda,\nu)=E+\lambda(Q-Q_0)+\nu(O-1),
\label{lm}
\end{equation}
where $\nu$ and $\lambda$ are Lagrange multipliers which enforce the
normalization and the desired value of $Q$, respectively. Taking the derivative of $L$
with respect to $c$, $\nu$ and $\lambda$ we get the familiar equations
\begin{equation}
{\partial L\over \partial c_{j}}=
\sum_k H_{jk} c_{k}+
\nu\sum_k O_{jk} c_{k}+
\lambda\sum_k Q_{jk} c_{k}=0,
\label{lagr}
\end{equation}
\begin{equation}
{\partial L\over \partial \nu}=O-1=0,
\end{equation}
\begin{equation}
{\partial L\over \partial \lambda}=Q-Q_0=0.
\label{conq}
\end{equation}
These equations determine the extremal values of $c$ and the values of
$\lambda$ and $\nu$. The actual calculation of these values, however, is not
simple. Without the constraint, Eq. \eqref{conq}, Eq. \eqref{lagr} is a
generalized eigenvalue problem and by solving it one obtains the
energy eigenvalues and orthogonal orbitals. With the constraint, 
Eq. \eqref{lagr} is not a solvable algebraic system
(except maybe if ${\hat H}$ and ${\hat Q}$ commute and have a common set of
eigenfunctions). One possible solution is to 
assume some value of $\lambda$ and try to iterate so that the
constraint is fulfilled.

Note, however, that the extremal value of $c$ is  not necessarily a
maximum or minimum of $L$. To ensure the minimum or maximum one has to
define  \cite{hancock}
\begin{equation}
L_{ij}={\partial L\over \partial c_{i} c_{j}}=
H_{ij}+ \nu A_{ij}+ \lambda Q_{ij},
\end{equation}
\begin{equation}
o_i=\sum_j O_{ij} c_{j},
\end{equation}
\begin{equation}
q_i=\sum_j Q_{ij} c_{j},
\end{equation}
and investigate the determinant
\begin{equation}
\det(e)=
\left\vert
\begin{matrix}
\ddots  &                           &        & \vdots   & \vdots \\ 
        & L_{ij}-e\delta_{ij} &        & o_i      & q_i    \\
        &                           & \ddots & \vdots & \vdots   \\
\hdots  & o_j                       & \hdots & 0      & 0        \\
\hdots  & q_j                       & \hdots & 0      & 0        \\
\end{matrix}
\right\vert=0.
\end{equation}
The expansion of $\det(e)$ is a polynomial of order $K-2$. The roots
of the polynomial are all positive if $E$ is a minimum at $c$, and are all negative if
$E$ is a maximum at $c$. Without the constraint, Eq. \eqref{conq}, this
polynomial can be used to prove the Ritz variational upper bounds
\cite{cohen}.

\revDK{Even if we would be able to determine $c$ using Eq. \eqref{lagr}, 
it is not guaranteed that the
energy would be minimized.} Section \ref{cio} details how the energy minimization and the
determination of the Lagrange multipliers can be done simultaneously. 

\subsection{The approach of Wu and Van Voorhis \label{WvV}}
Wu and Van Voorhis introduced an approach \cite{PhysRevA.72.024502}
in which Eq. \eqref{lagr}, the eigenvalue problem of the  Kohn-Sham
Hamiltonian,  is solved for a given $\lambda$ value. This $\lambda$ is
determined by minimizing
\begin{equation}
f(\lambda)=Q-Q_0=0
\end{equation}
by a root finding algorithm. One can, for example,  use a Newton iteration
\begin{equation}
\lambda^{(n)}=\lambda^{(n-1)}-\alpha 
{f(\lambda)\over f'(\lambda^{(n-1)})},
\label{WVl}
\end{equation}
where
\begin{equation}
f'(\lambda)={df(\lambda)\over d\lambda}
\end{equation}
and $\alpha$ is the step size. The derivative, $f'$, can be calculated
using perturbation theory \cite{PhysRevA.72.024502} or by finite
differencing. In the latter case,
\begin{equation}
f'(\lambda)={f(\lambda+\delta)-f(\lambda)\over \delta}
\end{equation}
has to be calculated for some small $\delta$ self-consistently. In
this approach, each energy minimizing self consistent loop has an
inner loop to find $\lambda$. 

\subsection{Simple iterative optimization\label{io}}
In this section we describe the iterative optimization of $\lambda$.
We drop the constraint of the normalization ($\nu=0$) 
and consider only the solution of Eqs. \eqref{lagr} and \eqref{conq}.
For a single orbital, the normalization will be enforced by normalizing
the wave function at each iteration, in the case of a set of orbitals, a
Gram-Schmidt orthogonalization step will be incorporated.

The simplest iterative solution is a steepest descent approach where
$c$ varies in the direction of the anti-gradient, 
\begin{equation}
c_{k}^{(n+1)}=c_{k}^{(n)}-x_0 {\partial L\over \partial c_{k}}=
c_{k}^{(n)}-x_0\left(\sum_j (H_{kj}+\lambda Q_{kj})c_{j}\right),
\label{gradstep}
\end{equation}
and $\lambda$ changes in the direction of the gradient, 
\begin{equation}
\lambda^{(n+1)}=\lambda^{(n)}+x_0 {\partial L\over \partial
\lambda}=\lambda^{(n)}+x_0\left(Q-Q_0\right).
\label{las}
\end{equation}
This is very closely related to the approach of Wu and Van Voorhis; Eq.
\eqref{gradstep} is a self-consistent minimization step and Eq.
\eqref{las} steers $\lambda$ toward the optimal value. The step in
Eq. \eqref{las} can be further improved by using $f'$ as in Eq.
\eqref{WVl} if $f'$ is readily available.

\subsection{Constrained iterative optimization\label{cio}}
Alternatively,  one can adjust $\lambda$ to fulfill the constraint. 
Unlike the simple update of $\lambda$ described in the previous
section, now we force the constraint on the interaction. 
Rewriting Eq. \eqref{gradstep} in matrix vector notation,
\begin{equation}
c^{(n+1)}=c^{(n)}-x_0(H+\lambda Q)c^{(n)},
\label{vec}
\end{equation}
the constraint can be written as
\begin{eqnarray}
Q_0&=&c^{(n+1)}Q c^{(n+1)} \\
&=&\left([I-x_0(H+\lambda Q)]c^{(n)}\right)Q
\left([I-x_0(H+\lambda Q)]c^{(n)}\right),
\nonumber
\end{eqnarray}
where for the left multiplication one uses the transpose of the
vector. Here we omit the transpose sign to simplify the notation. 
After dropping the terms that are quadratic in $x_0$ we can solve the
equation for $\lambda$,
\begin{equation}
\lambda={1\over  c^{(n)}Q^2 c^{(n)}}\left(
(c^{(n)}HQc^{(n)})-
{c^{(n)}Q c^{(n)}-Q_0\over 2 x_0}
\right),
\label{blam}
\end{equation}
and use this new $\lambda$ value in the iteration.  This expression
contains $Q^2$ and $HQ$ operators which are simple to evaluate in real
space approaches but could cause difficulties in other 
basis function representations. 

Alternatively \cite{Cusson1985,PhysRevC.32.172}, we can make an iteration for $\lambda$ in each step
adjusting it to improve the satisfaction of the constraint. We are
looking for the optimal $\delta\lambda$ so that the iteration
\begin{equation}
\lambda^{(n+1)}=\lambda^{(n)}+\delta \lambda
\end{equation}
converges to the optimal $\lambda$ value. 

The effect of an iterative step using $\delta\lambda Q$ alone is
\begin{equation}
c^{(n+1)}=c^{(n)}-x_0\delta\lambda Qc^{(n)}.
\end{equation}
The same procedure as above gives
\begin{equation}
\delta\lambda=
{c^{(n)}Q c^{(n)}-Q_0\over 2 x_0 c^{(n)}Q^2 c^{(n)}}
\label{blam1}
\end{equation}
as the optimal $\delta\lambda$ to enforce 
$Q_0=c^{(n+1)}Q c^{(n+1)}$. The same approach can also be used to
constrain the change of the expectation value of $Q$. In this case 
$\delta\lambda$ should be chosen as
\begin{equation}
\delta\lambda={c^{(n)}Q c^{(n)}-c^{(n+1)}Q c^{(n+1)}
\over 2 x_0 c^{(n)}Q^2 c^{(n)}}.
\label{blam2}
\end{equation}

In practice \cite{Cusson1985,PhysRevC.32.172}, the following update
algorithm proved to be efficient: 
\par\noindent
(1) Make an intermediate step
\begin{equation}
c^{(n+1/2)}=c^{(n)}-x_0(H+\lambda^{(n)} Q)c^{(n)}.
\end{equation}
\par\noindent
(2) Change $\lambda$ to the ideal value
\begin{eqnarray}
\lambda^{(n+1)}=\lambda^{(n)}&+&c_0{c^{(n+1/2)}Q c^{(n+1/2)}-c^{(n)}Q
c^{(n)}\over 2 x_0 c^{(n)}Q^2 c^{(n)}+d_0}\nonumber\\
&+&
{c^{(n)}Q c^{(n)}-Q_0\over 2 x_0 c^{(n)}Q^2 c^{(n)}+d_0}
\end{eqnarray}
(3) Advance iteration with the corrected $\lambda$
\par\noindent
\begin{equation}
c^{(n)}=c^{(n+1/2)}-x_0(\lambda^{(n+1)}-\lambda^{(n)})Q c^{(n+1/2)}.
\end{equation}
In step (2), both correction terms (Eqs. \eqref{blam1} and
\eqref{blam2}) are used: the first one reduces the change in the
expectation value and the second one adjusts the expectation value 
to its desired value. The numerical parameter, $c_0$, sets the relative
weight of the two terms. The second numerical constant, $d_0$, is a
parameter to compensate the neglected terms in deriving Eqs.
\eqref{blam1} and \eqref{blam2}. In step (3), $Q$ is multiplied by 
$(\lambda^{(n+1)}-\lambda^{(n)})$ and not by $\lambda^{(n+1)}$ because 
the $\lambda^{(n)}Q$ term has already acted on the wave function in
step (1). 

The extension of the above formalism for the many orbital case is
straightforward and the relevant equations are given in the main text. 

%

\end{document}